\documentclass[namedreferences]{solarphysics}
\usepackage[hyperref,optionalrh,solaromanenum]{spr-sola-addons} % For Solar Physics 
\usepackage{graphicx}                    % For eps figures, newer & more powerfull
\usepackage{color}                       % For color text: \color command
\usepackage{breakurl}                         % For breaking URLs easily trough lines
\newcommand{\eg}{{\it e.g.}}
\newcommand{\insitu}{{\it in-situ }}

\newcommand{\aap}{    {\it Astron. Astrophys.}}
\newcommand{\aaps}{   {\it Astron. Astrophys. Suppl.}}

\newcommand{\apj}{    {\it Astrophys. J.}}
\newcommand{\apjl}{   {\it Astrophys. J. Lett.}}

\newcommand{\grl}{    {\it Geophys. Res. Lett.}}

\newcommand{\jgr}{    {\it J. Geophys. Res.}}

\newcommand{\nat}{    {\it Nature}}

\newcommand{\pasj}{   {\it Pub. Astron. Soc. Japan}}

\newcommand{\solphys}{{\it Solar Phys.}}
 
\newcommand{\ssr}{    {\it Space Sci. Rev.}}

\begin{document}
%\linenumbers
\begin{article}

\begin{opening}

\title{Comparative Study of Microwave Polar Brightening, Coronal Holes, and Solar Wind Over the Solar Poles}

%%%%%%%%%%%%%%%%%%%%%%%%%%%%%%%%%%%%%%%%%%%%%%%%%%%
%% Authors Names
%
% \author[addressref={},corref,email={}]{\inits{}\fnm{}\lnm{}\orcid{}}

\author[addressref={aff1},corref,email={fujiki@isee.nagoya-u.ac.jp}]{\inits{K.}\fnm{Ken'ichi}~\lnm{Fujiki}}%\sep
\author[addressref={aff2},email={shibasaki.kiyoto@md.ccnw.ne.jp}]{\inits{K.}\fnm{Kiyoto}~\lnm{Shibasaki}}%\sep
\author[addressref={aff3,aff4},email={seiji.yashiro@nasa.gov}]{\inits{S.}\fnm{Seiji}~\lnm{Yashiro}}%\sep
\author[addressref={aff1},email={tokumaru@isee.nagoya-u.ac.jp}]{\inits{M.}\fnm{Munetoshi}~\lnm{Tokumaru}}%\sep
\author[addressref={aff1},email={k.iwai@isee.nagoya-u.ac.jp}]{\inits{K.}\fnm{Kazumasa}~\lnm{Iwai}}%\sep
\author[addressref={aff1},email={masuda@isee.nagoya-u.ac.jp}]{\inits{S.}\fnm{Satoshi}~\lnm{Masuda}}%\sep

\address[id=aff1]{Institute for Space-Earth Environmental Research, Nagoya University, F303(250), Furocho, Chikusa, Nagoya, 464-8601, Japan}
\address[id=aff2]{Solar Physics Research Inc., Matsushin, Kasugai, 486-0931, Japan}
\address[id=aff3]{Department of Physics, The Catholic University of America, Washington DC 20064, USA}
\address[id=aff4]{Code 671, NASA Goddard Space Flight Center, Greenbelt, Maryland, USA}

%%%%%%%%%%%%%%%%%%%%%%%%%%%%%%%%%%%%%%%%%%%%%%%%%%%
%% Runningheads
%
\runningauthor{Fujiki et al.}
\runningtitle{Microwave Polar Brightening, Coronal Holes, and Solar Wind}

%%%%%%%%%%%%%%%%%%%%%%%%%%%%%%%%%%%%%%%%%%%%%%%%%%%
%% Affilations 
%% id shold be the same with \author addressref value.
%\address[id={}]{}

%%%%%%%%%%%%%%%%%%%%%%%%%%%%%%%%%%%%%%%%%%%%%%%%%%%
%%% Abstract 
\begin{abstract}
We comparatively studied the long-term variation (1992--2017) in polar brightening observed with the Nobeyama Radioheliograph, the polar solar wind velocity with interplanetary scintillation observations at the Institute for Space-Earth Environmental Research, and the coronal hole distribution computed by potential field calculations of the solar corona using synoptic magnetogram data obtained at Kitt Peak National Solar Observatory. First, by comparing the solar wind velocity ($V$) and the brightness temperature ($T_b$) in the polar region, we found good correlation coefficients (CCs) between $V$ and $T_b$ in the polar regions, CC = 0.91 (0.83) for the northern (southern) polar region, and we obtained the $V$\,--$T_b$ relationship as $V$ =12.6 ($T_b$--10\,667)$^{1/2}+432$. 
We also confirmed that the CC of $V$\,--\,$T_b$ is higher than those of $V$\,--\,$B$ and $V$\,--\,$B/f$, where $B$ and $f$ are the polar magnetic field strength and magnetic flux expansion rate, respectively. 
These results indicate that $T_b$ is a more direct parameter than $B$ or $B/f$ for expressing solar wind velocity. Next, we analyzed the long-term variation of the polar brightening and its relation to the area of the polar coronal hole ($A$). As a result, we found that the polar brightening matches the probability distribution of the predicted coronal hole and that the CC between $T_b$ and $A$ is remarkably high, CC = 0.97. This result indicates that the polar brightening is strongly coupled to the size of the polar coronal hole. Therefore, the reasonable correlation of $V$\,--\,$T_b$ is explained by $V$\,--\,$A$. In addition, by considering the anti-correlation between $A$ and $f$ found in a previous study, we suggest that the $V$\,--\,$T_b$ relationship is another expression of the Wang--Sheeley relationship ($V$\,--\,$1/f$) in the polar regions. 
\end{abstract}

%%%%%%%%%%%%%%%%%%%%%%%%%%%%%%%%%%%%%%%%%%%%%%%%%%%
%% Keywords
%

\end{opening}
%-------------------------------------------------

%%%%%%%%%%%%%%%%%%%%%%%%%%%%%%%%%%%%%%%%%%%%%%%%%%%
%% Sections
%
% \section{}%\label{s:?} 
\section{Introduction}

The enhancement of brightness temperature around the solar pole is known as polar(-cap) brightening, observed at microwave to submillimetric wavelengths. The phenomenon was first reported by \inlinecite{Babin1976IzKry} with the observation of a raster scan of the solar disk at millimetric and centimetric wavelengths using the Crimea 22-m radio telescope. \inlinecite{Efanov1980IAUS} investigated a variation in polar brightening from 1968 to 1977 observed with the same radio telescope. They found an anti-correlation between solar cycle phases and polar brightenings, that is, polar brightenings were absent around the solar maximum and peaked in the solar minimum. \inlinecite{Kosugi1986PASJ} compared brightness temperatures in the polar and equatorial regions at 36 GHz using the Nobeyama 45-m telescope and suggested that the temperature and density structures of the upper chromosphere in the coronal hole differ from those outside the holes. 

The Nobeyama Radioheliograph (NoRH, \opencite{Nakajima1994IEEE}) is a solar-dedicated radio interferometer, and it has been providing well-calibrated and highly resolved full-disk images every day since mid-1992. Investigations of long-term variations in the polar brightening over a solar cycle can be realized only by using the NoRH.

\inlinecite{Gelfreikh2002A&A-1}	 produced a microwave butterfly map at 17 GHz for 1992 to 2001 and compared the latitudinal polar faculae distribution over the same period. They found a global coincidence in the latitudinal distribution of these phenomena. However, not all microwave enhancements were found to correspond to the polar faculae with a one-to-one correspondence. \inlinecite{Gopa2000JApA}  investigated 71 equatorial coronal holes that had been observed between 1996 and 1999 using NoRH, satellites, and ground-based observations. They found a positive correlation between the microwave brightness temperature and the peak magnitude of the photospheric magnetic field of the coronal holes. 
\inlinecite{Gopals2012ApJ} identified a strong correlation between the polar brightening and the polar magnetic field observed at the Kitt Peak National Solar Observatory (KP/NSO). They identified a north–-south asymmetry of the CC in both polar regions during the 22/23 and 23/24 minima, that is, the CC in the southern polar region was found to exhibit a higher CC (CC = 0.86 and 0.71) than that in the northern polar region (CC = 0.64 and 0.54) during each solar minimum. Moreover, the CCs in both polar regions became weaker in the 23/24 minimum when the solar activity weakened. The strong correlation (CC = 0.86) was also observed by \inlinecite{Shibasaki2013PASJ} based on the polar magnetic field observed at the Wilcox Solar Observatory (WSO). He also reported on microwave enhancements of 6\,--\,16 \% from the quiet solar-disk level and their solar-cycle dependence through nearly two solar cycles.
The strong correlation indicates an association between the polar brightening and the polar coronal hole, as is mentioned in several other studies (\opencite{Selhorst2003A&A, Nitta2014ApJL, Kim2017JKAS}).

The association between the solar wind velocity and coronal holes has been discussed ever since the discovery of coronal holes, which were thought to be related to the recurrent fast solar winds (\opencite{Krieger1973SoPh}). A direct comparison between the properties of the low-latitude coronal holes at the central meridian and the velocities of the solar winds identified as the cause of the coronal holes revealed that coronal holes are the origin of not only fast but also slow solar winds (\opencite{Nolte1976SoPh, Akiyama2013PASJ}). 

 Recently, because of the importance of the solar wind velocity prediction for space weather forecasts, the empirical modeling of the solar wind velocity with various parameters was investigated. The size of the coronal hole is one of the important parameters in solar wind prediction because the size of coronal holes is easily detected by real-time imaging observations in soft X-ray and EUV emissions of the Sun (\opencite{Rotter2015SoPh, Reiss2016SpWea}).
 
 The polar coronal holes are also known as the origin of the fast solar wind. 
Direct measurements of the solar wind around the solar poles have been made by only the Ulysses spacecraft (\opencite{Bame1992A&AS}), which escaped from the ecliptic plane using a gravity assist from Jupiter. The global structure of the solar wind was revealed by Ulysses in a wide latitude range (--82$^{\circ}$ to 82$^{\circ}$) including three polar-pass missions (\opencite{McComas2008GRL}).
During a solar minimum, the global solar wind structure is highly stratified in latitude (\opencite{McComas2000JGR}). Slow solar wind is confined to a narrow region around the solar equator. At higher latitudes, the solar wind velocity increases from 400 to 750 km/s (\opencite{McComas1998JGR}). The fast solar wind in the polar region possesses an extremely constant feature in various parameters (\opencite{Phillips1995Sci, McComas2000JGR}). A similar solar wind structure had been obtained in interplanetary scintillation (IPS) observations in an earlier study (\opencite{Kojima1990SSRv}).
In contrast, the polar solar wind in a solar maximum is highly variable (\opencite{McComas2002GeoRL}). The fast solar wind disappears as the polar coronal hole shrinks and disappears. Just after the solar maximum, the polar coronal hole reappears with an opposite polarity and plays a role in the origin of the fast solar wind. The centroid of the polar coronal hole in this period does not locate the solar poles (\opencite{Fujiki2016ApJL}). In this case, the polar region is not covered uniformly by the fast solar wind, because of the deviation of the polar coronal hole from the solar pole. Ulysses measured the intermittent fast solar wind around the north polar region in 2001 (\opencite{McComas2002GeoRL}).

IPS observations (\opencite{Hewish1964Nature}) are one of the useful tools for investigating the global structure of the solar wind. Sketches of the global solar wind structure are available by accumulating IPS data for a relatively short period (typically one Carrington rotation; CR).
\inlinecite{Kojima1999JGR} identified a low-latitude coronal hole near the active region as the origin of the slow solar wind observed near the heliospheric current sheet by using data of IPS observations, \insitu measurements, and the Potential Field Source Surface (PFSS) model. \inlinecite{Ohmi2001JGR} analyzed the polar coronal hole immediately before the solar maximum and found that the fast solar wind originating from the shrinking polar coronal hole temporarily turns into the slow solar wind. After the disappearance of the polar coronal hole, the polar region in interplanetary space is filled by the slow solar wind originating from the coronal hole located at lower latitudes. \inlinecite{Tokumaru2017SoPh} investigated the relationship between coronal hole size and solar wind velocity by using the IPS data and calculated the predicted coronal hole size from the PFSS model from 1995 to 2011. They found a linear or quadratic correlation between those parameters.

\inlinecite{Akiyama2013PASJ} investigated 21 equatorial holes that appeared between 1996 and 2005. They analyzed the correlation between the solar wind measured by the Advanced Composition Explorer (ACE) spacecraft and various parameters affecting coronal holes, including microwave enhancement derived from NoRH, the Extreme-ultraviolet Imaging Telescope (EIT), and the  Michelson Doppler imager (MDI) on board the Solar and Heliospheric Observatory (SOHO). They identified several useful relationships for predicting the solar wind velocity. However, the polar coronal hole has not yet been investigated in the same degree of detail because it is difficult to continuously monitor the solar wind originating from the polar coronal hole by \insitu measurements.    
 
\inlinecite{Gopals2018JASTP} analyzed the polar brightening and polar solar wind measured by the Solar Wind Over the Poles of the Sun (SWOOPS) instrument aboard the Ulysses spacecraft during polar-pass intervals and found that the polar brightening is significantly correlated with the velocity of the fast solar wind. Because the three polar-pass intervals were limited in two solar minima and one solar maximum, the relationship has not yet been confirmed through a full solar cycle. By comparing long-term variations obtained by NoRH and IPS observations, we expect that the solar cycle dependence of the relationship will be revealed statistically. 

In this study, we investigate the relationship between the polar brightening observed with the NoRH, the coronal hole predicted by the PFSS model, and the polar solar wind with IPS observations. In section \ref{sec:but}, we describe the various types of butterfly maps. We demonstrate the temporal variation of the above parameters and their correlations in section \ref{sec:results}. Then, we discuss and summarize our results in sections \ref{sec:discussion} and \ref{sec:summary}, respectively.

\section{Various Butterfly Maps \label{sec:but}}

In this section, we introduce various `butterfly' maps. All maps have the same format with a temporal resolution of one CR (horizontal axis) and a regularly re-gridded latitude from $-89.5^{\circ}$ to $89.5^{\circ}$ with a one-degree grid (vertical axis).

\subsection{Radio Butterfly Map}
\begin{figure} 
	\centerline{\includegraphics[width=0.8\textwidth,clip=]{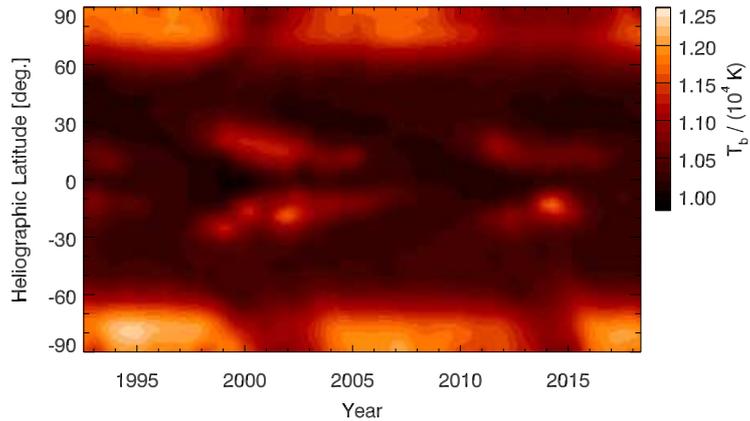}}
	\caption{Radio butterfly map obtained with NoRH.}\label{fig:norh_bfly}
\end{figure}

NoRH has been observing the upper chromosphere of the Sun every day at a frequency of 17 GHz since mid-July 1992. At this frequency, the brightness temperature of the quiet solar disk has been determined by a single-dish observation as 10\,000 K (\opencite{Zirin1991ApJ}), which is used for intensity calibration in synthesis imaging. We have constructed a radio synoptic map and radio butterfly map from July 1992 (CR 1858) to June 2018 (CR 2204) with the same method used by \inlinecite{Shibasaki2013PASJ}. Because the solar rotation axis tilts 7.25$^\circ$ with respect to the ecliptic plane (solar $B_0$ angle), the polar region of the radio butterfly map is affected by seasonal variations. In addition, the synthesized beam, which depends on the solar altitude, causes another seasonal variation in spatial resolution. To remove these effects on the butterfly map, we use a 13-CR running-mean and a Gaussian filter with a width of three pixels. Figure \ref{fig:norh_bfly} presents the radio butterfly map for the above interval. 

%fig1
%%%%%%%%%%%%%%%%%%%%%%%%%%%%%%%%%%%%%%%%%%%%%%%%%%%%%%
% Figure 
%%%%%%%%%%%%%%%%%%%%%%%%%%%%%%%%%%%%%%%%%%%%%%%%%%%%%%
\subsection{Solar Wind Butterfly Map \label{sec:vmap} }
\begin{figure} 
	\centerline{\includegraphics[width=0.8\textwidth,clip=]{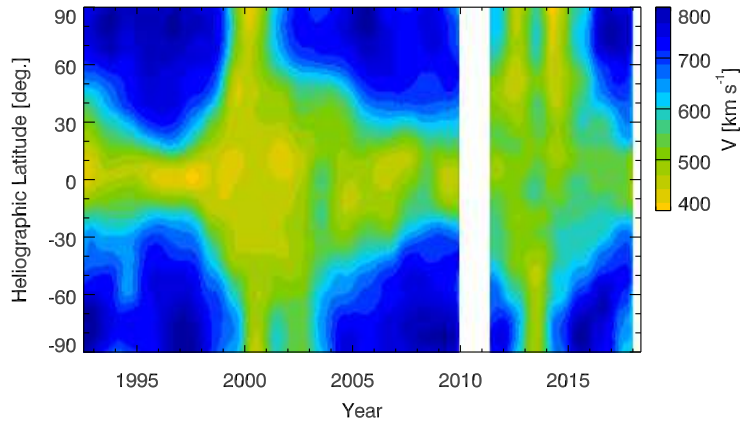}}
	\caption{Solar wind butterfly map obtained with the IPS observation. During the white gap, the IPS observation stopped because of an upgrade to the multi-site IPS system. }\label{fig:ips_bfly}
\end{figure}

Multi-site IPS observations at a frequency of 327 MHz have been carried out at the Institute for Space-Earth Environmental Research (ISEE, formerly Solar-Terrestrial Environment Laboratory), Nagoya University, Japan since the early 1980s (\opencite{Kojima1990SSRv, Tokumaru2010JGR}). In general, the IPS signal is the result of a weighted integration of solar wind velocity and density irregularity from observer to radio source (\eg\ see \opencite{Kojima1998JGR}), which is an obstacle to the accurate estimation of the solar wind structure. To reduce the integration effects, we introduce several types of IPS tomographic analyses (\opencite{Kojima1998JGR, Fujiki2003AnGeo}). In this study, we use time-sequence tomography (\opencite{Fujiki2003AnGeo}) to produce a solar-wind butterfly map from CR 1858 to CR 2198. The multi-site IPS observation is stopped during the winter because of snowfall in the mountainous site every year and is also interrupted because of unexpected observation system issues. These interruptions of observation produce an undesirable data gap on the solar wind velocity map. To fill the data gaps, we applied the same temporal filter as the NoRH butterfly map. Figure \ref{fig:ips_bfly} displays the solar wind butterfly map obtained by the above procedure. We remove the data in 2010 from the solar wind butterfly map because the IPS observation was almost fully unavailable in the year because of an upgrade to the multi-site IPS observation system. 
%%%%%%%%%%%%%%%%%%%%%%%%%%%%%%%%%%%%%%%%%%%%%%%%%%%%%%
%fig2
%figure
%%%%%%%%%%%%%%%%%%%%%%%%%%%%%%%%%%%%%%%%%%%%%%%%%%%%%%
\subsection{Magnetic Field and Coronal Hole Butterfly Map}
\begin{figure} 
	\centerline{\includegraphics[width=0.8\textwidth,clip=]{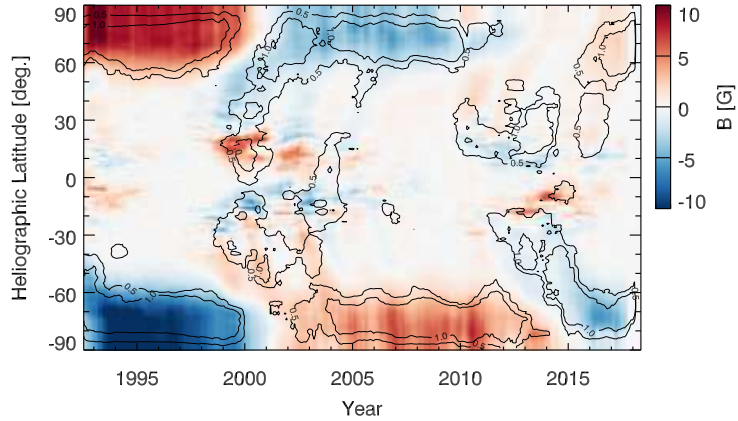}}
	\caption{Magnetic butterfly map (color) and probability distribution of open magnetic field region (P; contour). The red and blue regions in the map denote positive and negative polarities, respectively. The contour levels are 0.5\% and 1.0\%.}\label{fig:kpnso_bfly}
\end{figure}

In order to obtain the photospheric magnetic field and coronal hole distribution, we use the solar magnetic field data observed at KP/NSO and calculated the coronal magnetic field configuration using the PFSS model (\opencite{Schatten1969SoPh, Altschuler1969SoPh}). We use the PFSS analysis code developed by \inlinecite{Hakamada1995SoPh} and set the height of the source surface to 2.5 solar radii. In each CR from CR 1858 to CR 2198, we identify footpoints of the magnetic field open to interplanetary space by the PFSS extrapolation. Then, we integrate an area of the open magnetic field region in units of km$^2$ at each latitude bin of one degree ($A_{\theta}$) and the total area ($A$). We use a parameter, $P=A_{\theta}/A$, as the latitudinal probability distribution of the open magnetic field region, which reflects the coronal hole distribution in each CR. By arranging the probability distribution, we obtain a coronal hole butterfly map. Figure \ref{fig:kpnso_bfly} presents the probability distribution contour overlaid on the KP/NSO magnetic butterfly map smoothed by the 13-CR running-mean. The counter levels denote the probability of 0.5\% and 1.0\%, which clearly match the variation in coronal magnetic field observed on the magnetic butterfly map. Further, note that open footpoints are concentrated in a relatively intense unipolar region at mid-latitudes, especially around the solar maxima. 
%%%%%%%%%%%%%%%%%%%%%%%%%%%%%%%%%%%%%%%%%%%%%%%%%%%%%%
%fig3
%figure
%%%%%%%%%%%%%%%%%%%%%%%%%%%%%%%%%%%%%%%%%%%%%%%%%%%%%%
\section{Results \label{sec:results}}
 \subsection{Latitude Profiles of Brightness Temperature}
\begin{figure} 
	\centerline{\includegraphics[width=0.8\textwidth,clip=]{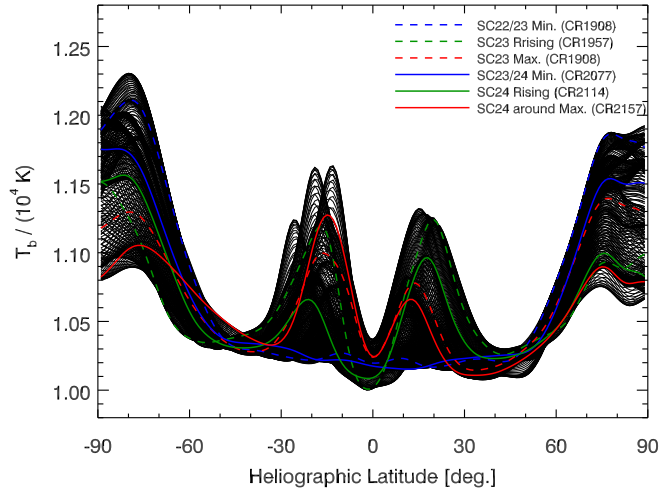}}
	\caption{Latitude profiles of $T_b$ for each CR. Solid and dashed curves plotted in color specify typical solar cycle phases. The legend details each curve.}\label{fig:lat_prof}
\end{figure}

First, we define a start latitude of the polar brightening to analyze the polar region. Figure \ref{fig:lat_prof} demonstrates the latitude profile of brightness temperature in units of Kelvin for all CRs specified with typical solar cycle phases in color (solar minimum, rising phase, and solar maximum). 

Around the equator, $0^{\circ} < |\theta| < 10^{\circ}$, $T_b$ exhibits a small variability except in the rising phase. The small dip ($\sim -200$ K) may be attributed to the effect of a dark filament because the activity of dark filaments increases at the beginning of the rising phase (\eg,\,see \opencite{Gopals2012ApJ}). The profiles are highly variable between $10^{\circ} < |\theta| < 40^{\circ}$ depending on the solar cycle phases because of the active region distribution. The local peaks coincide with the centroid of the butterfly wing. The mid-latitude region, $40^{\circ} < |\theta| < 50^{\circ}$, is not disturbed by $T_b$ enhancements, because of active regions and polar brightenings in the entire period. We derive the north--south asymmetry of $T_b$ in this region, and we obtain 10\,222 $\pm$ 49 K for the north and 10\,338 $\pm$ 41 K for the south. The difference in brightness temperature in the mid-latitude region is 154 K.
$T_b$ starts to increase rapidly at $50^\circ<|\theta|<60^\circ$ in the entire period. In a higher latitude region, $|\theta|>75^\circ$, $T_b$ reaches a peak and decreases at the solar poles in most of the period. The decrease in $T_b$ near the solar pole is caused by a contamination of radiation from outside the solar limb because of the finite beam size of NoRH (10\,--\,15 arcseconds) in synthesis imaging. $T_b$ is biased negatively at the limb observation, and this bias affects $T_b$ around the solar pole ($|\theta|>\sim82^\circ$) because of the seasonal variation of solar $B_0$ angle. As a result, the $T_b$ around the solar pole is underestimated, and it gives a lower limit of the polar brightening.

Based on the above analysis, we determine the latitude, $|\theta| = 60^{\circ}$, as the starting latitude of the polar brightening. In subsequent sections, we use the terminology, ``polar region,'' for regions with higher latitudes than $60^{\circ}$. Note that all the parameters for the correlation analyses are averaged values over $\theta>60^\circ$ with a weighting function that varies with the latitude, $\cos \theta$. This reduces the effect of the aforementioned contamination.
%%%%%%%%%%%%%%%%%%%%%%%%%%%%%%%%%%%%%%%%%%%%%%%%%%%%%%
%fig4
%figure
%%%%%%%%%%%%%%%%%%%%%%%%%%%%%%%%%%%%%%%%%%%%%%%%%%%%%%
 \subsection{Correlation between Polar Brightening and Polar Magnetic Field Strength}

\begin{figure} 
	\centerline{\includegraphics[width=0.8\textwidth,clip=]{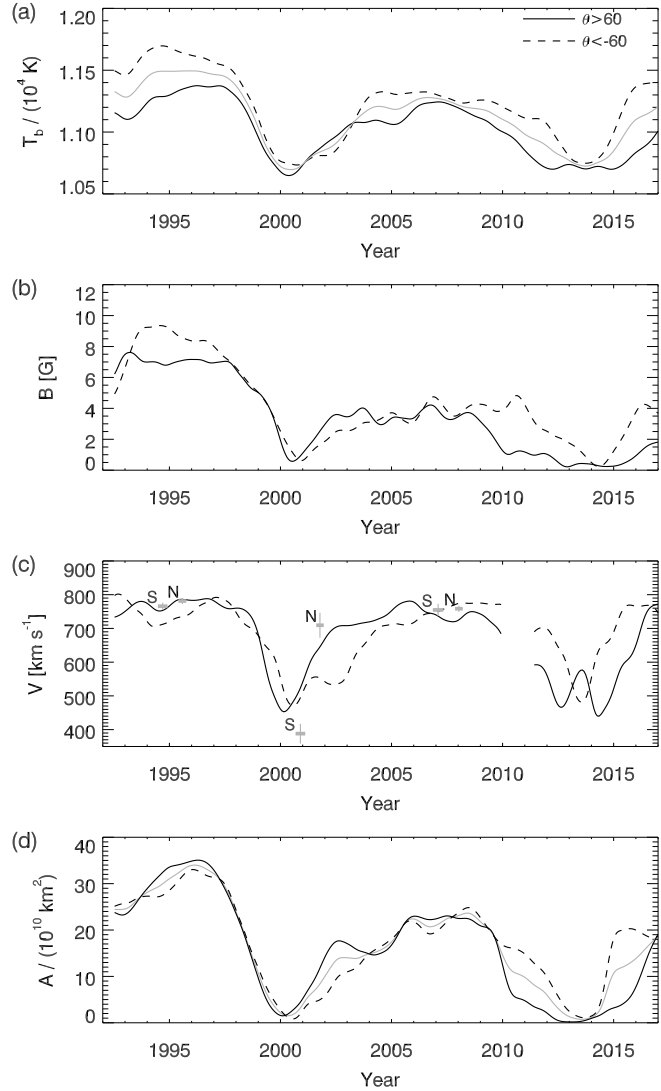}}
	\caption{(a) Temporal variation of brightness temperature ($T_b$ [K]). The solid and dashed curves represent those in the northern and southern polar region respectively. The gray curve represents the average. (b) Temporal variation of magnetic field strength ($B$ [G]). (c) Temporal variation of solar wind velocity ($V$ [km/s]). The horizontal thick gray lines indicate the average proton velocity measured with SWOOPS/Ulysses obtained during three polar passes. N and S denote the northern and southern polar region, respectively. The vertical lines represent the one-sigma level. (d) Temporal variation of computed coronal hole area ($A$ [km$^2$]) in the polar region. The gray curves represent the average. Each data point in all panels has been obtained by averaging over $\theta > 60^{\circ}$ with a weight of $\cos \theta$.}\label{fig:time_prof}
\end{figure}

Here, we examine the consistency of our results with those of previous studies, applying different datasets and temporal filters to the data. 
Figures \ref{fig:time_prof} (a) and (b) show the temporal variation of $T_b$ and $B$, respectively. $T_b$ shows the periodic variation, peaking around the solar minima and reaching the lowest around the solar maxima, and a similar variation is found in $B$. However, small fluctuations do not coincide with each other. Variations during the last solar maximum north--south asymmetry are remarkable. $T_b$ and $B$ recover quickly within two years in the south polar region. In contrast, darker $T_b$ and weak $B$ persist in the north polar region. This north--south asymmetry is consistent with the coronal hole activity in the 24th solar maximum. 
%%%%%%%%%%%%%%%%%%%%%%%%%%%%%%%%%%%%%%%%%%%%%%%%%%%%%%
%fig5
%figure
%%%%%%%%%%%%%%%%%%%%%%%%%%%%%%%%%%%%%%%%%%%%%%%%%%%%%%

\begin{figure} 
	\centerline{\includegraphics[width=0.6\textwidth,clip=]{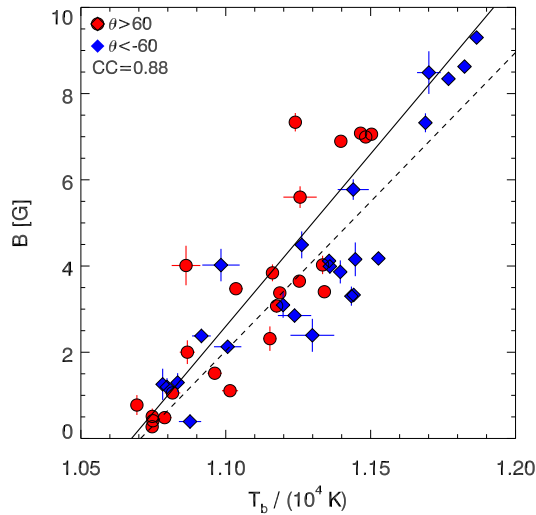}}
	\caption{Correlation plot between $T_b$ and $B$. The data points have been obtained by averaging the temporal variation in each year. The red and blue points denote the northern and southern polar regions, respectively. The error bars denote the one-sigma level. The red and blue lines represent regressions for data points of the same color. The CC is 0.88.}\label{fig:B_Pb}
\end{figure}

Figure \ref{fig:B_Pb} shows a correlation plot between yearly averaged $T_b$ and $B$. $T_b$ is linearly proportional to $B$; however, data points in the intermediate range are relatively scattered. The CC between these two parameters is high (CC = 0.88). 

\inlinecite{Gopals2018JASTP} obtained a lower CC (0.72) from the NoRH and KP/NSO data than that obtained in the present study.  This difference may be caused by the temporal filter. Since the KP/NSO data were not 13-CR running-averaged in their study, the scatter of the data points was relatively large. However, it is difficult to compare the CCs directly because the number of data points differs. The amount of data used in their study was much greater ($N=678$) than that used in the present study ($N = 26$). The Pearson critical correlation coefficients are 0.132 (\opencite{Gopals2018JASTP}) and 0.634 (this study) for a significance level of $p = 5\times10^{-4}$, indicating that both results are statistically significant.
 On the other hand, \inlinecite{Shibasaki2013PASJ} obtained a CC (0.86) for the NoRH and WSO data that was comparable to that which we obtained. The WSO magnetogram has a much lower resolution than that of KP/NSO. In addition, \inlinecite{Shibasaki2013PASJ} applied a temporal filter (20 nHz) to the magnetic butterfly map. These factors supressed the scattering of the data points and yielded a higher CC. In the present study, we used a temporal filter that was almost the same as that used by \cite{Shibasaki2013PASJ}.

%
%%%%%%%%%%%%%%%%%%%%%%%%%%%%%%%%%%%%%%%%%%%%%%%%%%%%%%
%fig6
%figure
%%%%%%%%%%%%%%%%%%%%%%%%%%%%%%%%%%%%%%%%%%%%%%%%%%%%%%
 \subsection{Correlation between Polar Brightening and Polar Solar Wind}

Figure \ref{fig:time_prof} (c) shows the temporal variation in polar solar wind in units of km/s sampled from the solar wind butterfly map. The velocity of the polar solar wind ($|\theta|>60^{\circ}$ in interplanetary space) exceeds 700 km/s during most of the period except for two solar maxima, in which large coronal holes in both polar regions are stationary. In this period, the north--south asymmetry is small.

In contrast, the north--south asymmetry is remarkable in the 23rd and 24th solar maxima. Just before the 23rd solar maximum, the fast solar wind in both polar regions turns into slow solar wind almost simultaneously in 1999. Then, the northern polar fast solar wind recovers in 2002 simultaneously; however, the southern polar fast wind disappears again in 2003. During the 24th maximum, the fast solar wind in the northern polar region disappears in 2012, which is a precedence of one year over that in the southern polar wind. The recovery of the fast solar wind in the southern polar region is a precedence of more than one year compared to that in the northern polar region. The northern polar fast wind recovers in 2013 and disappears again in 2014. The variation in the northern polar region resembles that of the southern polar region in the 23rd maximum. The north--south asymmetry in 1985\,--\,2013 observed with the IPS has been discussed in \inlinecite{Tokumaru2015JGR}.

Mean proton velocities measured by the SWOOPS/Ulysses in 1994\,--\,1995 (first polar pass), 2000\,--\,2001 (second polar pass), and 2007\,--\,2008 (third polar pass) have been plotted by thick gray lines in figure \ref{fig:time_prof} (c). N and S denote the northern and southern polar region, respectively. The velocity of the polar solar wind coincides well in the first and third polar passes during two solar minima. In contrast, large differences are found in the second polar pass during the solar maximum.
\inlinecite{Fujiki2003AnGeo} compared the IPS and Ulysses data in 1999\,--\,2000 and found a good correspondence in both data sets even in the solar maximum when the solar wind structure is highly complex. Therefore, this large difference is caused by a smoothing effect of the solar wind butterfly map.

\begin{figure} 
	\centerline{\includegraphics[width=0.6\textwidth,clip=]{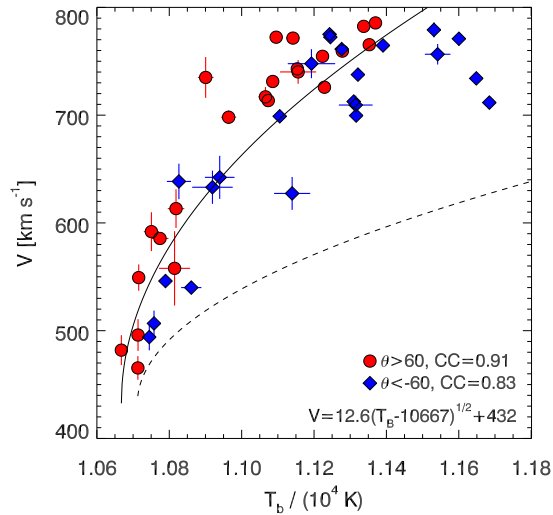}}
	\caption{
		Correlation plot between $T_b$ and $V$. The color and error-bars are the same as in figure \ref{fig:B_Pb}. The CCs are 0.91 (0.83) for the northern (southern) polar region. The solid curve represents the regression of the relationship between $V$ and $T_b^{1/2}$, and the dashed curve is the regression obtained by \inlinecite{Tokumaru2017SoPh}, which has been converted to a function of $T_b$ using equation (\ref{eq:PB_A}) in section \ref{sec:discussion}. 
	}\label{fig:V_Pb}
\end{figure}

Figure \ref{fig:V_Pb} shows a correlation plot between $V$ and $T_b$. The correlation plot of the northern polar region (red) and southern polar region (blue) appears to be shifted slightly. 

This result is consistent with that obtained by \inlinecite{Gopals2018JASTP}. 

In addition, the slope of the plot becomes moderate or flat at higher $T_b$. The CCs are 0.91 and 0.83 for the northern and southern polar regions.

Here, it is important to note that the IPS observation in the southern polar region is less accurate than that in the northern polar region. This is an inevitable problem because of the geographical conditions of the ISEE-IPS observatory in Japan. The number of IPS radio sources, which affects the accuracy of the IPS tomography, in the southern region of the Sun is always smaller than that in the northern region of the Sun. However, it does not cause a serious problem, because yearly averaged data are used for all analyses in this study. The temporal averaging covers the deficiency of observations in south polar region.
%%%%%%%%%%%%%%%%%%%%%%%%%%%%%%%%%%%%%%%%%%%%%%%%%%%%%%
%fig7
% figure here
%%%%%%%%%%%%%%%%%%%%%%%%%%%%%%%%%%%%%%%%%%%%%%%%%%%%%%
 \section{Discussion \label{sec:discussion}}
In figure \ref{fig:lat_prof}, the flat disk component exhibits a north--south asymmetry, that is, $T_b$ is systematically higher in the southern hemisphere than in the northern hemisphere through solar cycles. It is clear from the lower envelope of the latitude profiles that the $\mathrm{d}T_b/\mathrm{d}\theta$ is a constant. By using the north--south asymmetry at $40^{\circ} < |\theta| < 50^{\circ}$, we obtain the latitudinal profile expressed as $T_b = -1.29 \theta +10\,280$ K. This asymmetry causes a difference in $T_b$ of 154 K at $|\theta| = 60^{\circ}$ and 233 K at the solar poles if we extrapolate linearly to the solar poles. It is difficult to explain the constant decline of the disk brightness temperature only with the dark filament distribution, because $\mathrm{d}T_b/d\theta$ should be reversed in both hemispheres by the dark filament. This strange feature, even seen in solar minima, is quite interesting and worth investigating in the future. 

\begin{figure} 
	\centerline{\includegraphics[width=0.6\textwidth,clip=]{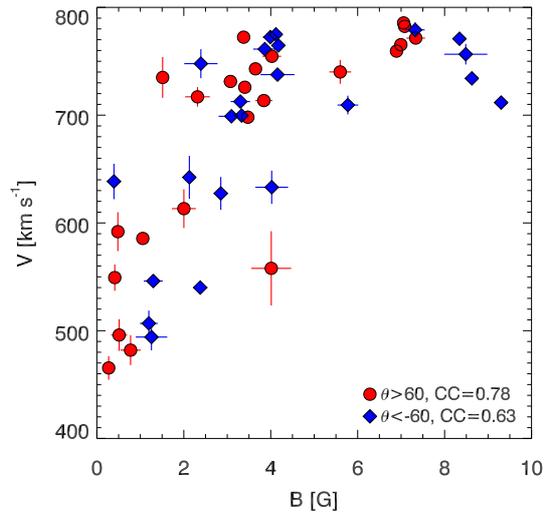}}
	\caption{Correlation plot between $B$ and $V$. The color and error-bars are the same as in figure \ref{fig:B_Pb}. The CCs are 0.77 (0.61) for the northern (southern) polar region.}\label{fig:V_B}
\end{figure}

Good correlations are obtained for $V$\,--\,$T_b$ and $T_b$\,--\,$B$; therefore, we investigate the correlation of $V$\,--\,$B$ to confirm which parameter, $T_b$ or $B$, is suitable for the expression of the solar wind velocity. 
Figure \ref{fig:V_B} shows a correlation plot between $B$ and $V$. The CCs are 0.78 and 0.63 for the northern polar region (red) and southern polar region (blue), respectively. These CCs are high but sufficiently smaller than the $V$\,--\,$T_b$ relationship, which indicates that the $T_b$ is a more direct parameter for the solar wind acceleration in the polar region. 
%%%%%%%%%%%%%%%%%%%%%%%%%%%%%%%%%%%%%%%%%%%%%%%%%%%%%%
%fig8
%%%%%%%%%%%%%%%%%%%%%%%%%%%%%%%%%%%%%%%%%%%%%%%%%%%%%%

\begin{figure} 
 	\centerline{\includegraphics[width=0.8\textwidth,clip=]{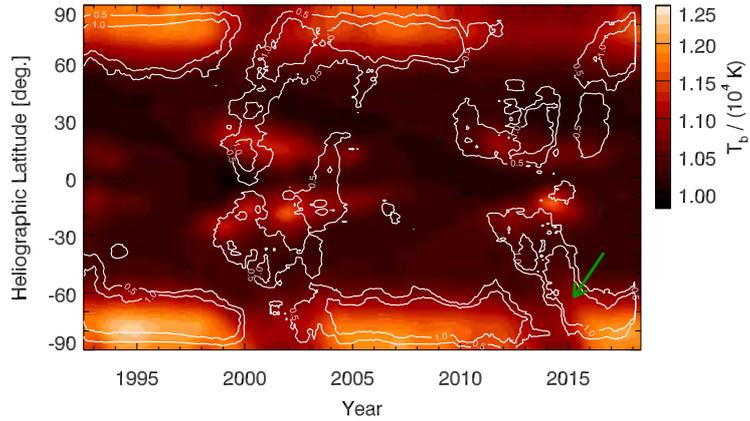}}
 	\caption{Probability distribution of open magnetic field region overlaid on radio butterfly map. The contour levels are the same as in figure \ref{fig:kpnso_bfly}} \label{fig:norh_CH_bfly}
 \end{figure}

 Many works have pointed out the relationship between the polar brightening and coronal hole (\opencite{Selhorst2003A&A, Shibasaki2013PASJ, Kim2017JKAS}). We also investigate the relationship between the coronal holes predicted by the PFSS model in figure \ref{fig:norh_CH_bfly}. The contour levels, $P$ = 0.5\% and 1.0\%, covering the polar regions in the entire period, except around a solar maximum, show a good correspondence with the polar brightening. Additionally, in some regions with a high coronal-hole probability along the meridional flow, the enhancement of brightness temperature is identified in the contour. The most prominent region is specified by a green arrow. This feature suggests that the polar brightening is produced by not only the line-of-sight geometry but also an additional mechanism such as a different condition of the solar atmosphere in a coronal hole. Brightness enhancements were found in the elephant trunk during the Whole Sun Month campaign (August 10 to September 9, 1996). In addition, microwave enhancements are associated with unipolar magnetic flux elements in the magnetic network (\opencite{Gopals1999JGR}). 
 The brightness enhancement in the mid-latitude region detected on the radio butterfly map indicates that a relatively large coronal hole with a long lifetime appeared at the southern mid-latitude in 2014.
 
 We do not discuss the mechanism of the microwave enhancement in this study, however, 
 
  comprehensive studies of the coronal hole using recent sophisticated data may provide key information on the mechanism of microwave brightness enhancements in a coronal hole
  
   coupling with a foresighted arguments in \inlinecite{Gopa1999AIPC}.

%%%%%%%%%%%%%%%%%%%%%%%%%%%%%%%%%%%%%%%%%%%%%%%%%%%%%%
 %fig9
%%%%%%%%%%%%%%%%%%%%%%%%%%%%%%%%%%%%%%%%%%%%%%%%%%%%%%
 
\begin{figure} 
 	\centerline{\includegraphics[width=1.0\textwidth,clip=]{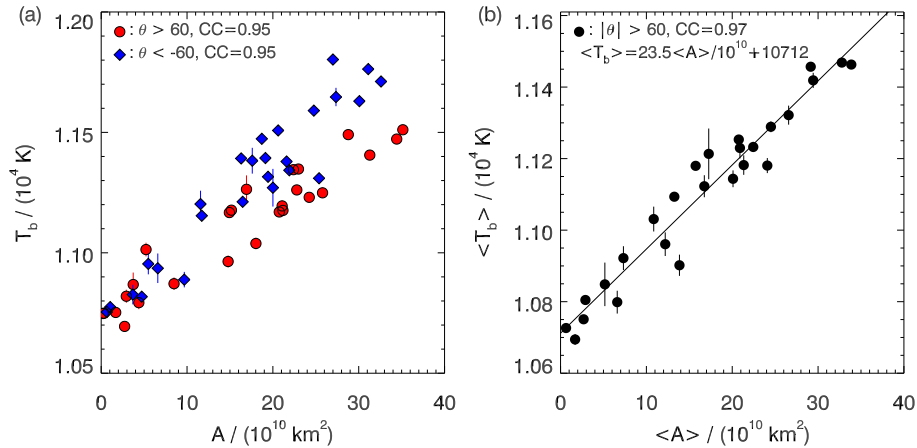}}
 	\caption{(a) Correlation plot between $A$ and $T_b$. The CCs in the two parameters are both 0.95. (b) Correlation plot between $\left\langle A\right\rangle$ and $\left\langle T_b\right\rangle$.  Brackets indicate the average of the parameters for both polar regions. The red line shows a regression of the $\left\langle A\right\rangle$\,--\,$\left\langle T_b\right\rangle$ relationship. The CC is 0.97.}\label{fig:A_Pb}
 \end{figure}

 Figure \ref{fig:time_prof} (d) demonstrates the temporal variation of the physical area of the polar coronal hole ($A$). During solar maximum when the polar brightening becomes minimum $T_b$, no coronal hole is found in the polar region. 
 Figure \ref{fig:A_Pb} shows correlation plots between $T_b$ and $A$. Figure \ref{fig:A_Pb} (a) shows a systematic asymmetry in the northern and southern polar regions. The CCs are 0.95 for both polar regions. Surprisingly, if we plot a correlation diagram between the 
 
 averaged values, $\left\langle T_b\right\rangle $ and $\left\langle A\right\rangle $ 
 
 in both polar regions, the CC improves to 0.97. This should be due to the cancellation of the north--south asymmetry of the brightness temperature, $\Delta T_b=116$ K in 40$^{\circ}$\,--\,50$^{\circ}$.
 From the data points in figure \ref{fig:A_Pb} (b), we obtain the regression line as
 \begin{equation}\label{eq:PB_A}
 \left\langle T_b\right\rangle  = 23.5 \frac{\left\langle A\right\rangle }{10^{10}} +10\,712.
 \end{equation}
 Even when the polar coronal hole disappears in a solar maximum, $\left\langle T_b\right\rangle$ has a finite excess of brightness temperature, about 700 K, from the quiet disk level (10\,000 K). We note that the 7\% enhancement of polar brightening reasonably matches the 10\% enhancement of limb brightening at mid-latitudes obtained by \inlinecite{Selhorst2003A&A}. As shown in figure \ref{fig:lat_prof}, the mid-latitude region is less affected by sunspot activity through a solar cycle. Therefore, the enhancement is the limb brightening of the quiet region, and the polar brightening is linearly proportional to the size of the polar coronal hole with an offset of about 700 K. Actually, the result is consistent with \inlinecite{Gopals2016ApJ}, who found that the polar brightening from the quiet sun level dropped to about 700 K in 2014 under the condition of the longer-lasting zero-B around the northern polar region. 
 %%%%%%%%%%%%%%%%%%%%%%%%%%%%%%%%%%%%%%%%%%%%%%%%%%%%%%
 %fig10
%%%%%%%%%%%%%%%%%%%%%%%%%%%%%%%%%%%%%%%%%%%%%%%%%%%%%%
 \inlinecite{Tokumaru2017SoPh} found a good correlation (CC = 0.66) between $V$ and $A$ by statistical analysis using IPS and magnetogram observations. They found a linear or quadratic relationship. The quadratic relationship gives a slightly better correlation and the following formula:
 \begin{equation}\label{eq:tok}
 % V = 30.2 (A/10^{10})^{1/2} + 433\ \  [\mathrm{km/s}]. 
 V = 30.2 \left( \frac{A}{10^{10}} \right)^{1/2} + 433. 
 \end{equation}
 The dashed curve in figure \ref{fig:V_Pb} represents the $V$\,--\,$A^{1/2}$ relationship of equation (\ref{eq:tok}), which has been transformed as a function of $T_b$ by equation (\ref{eq:PB_A}).
 The solid curve in figure \ref{fig:V_Pb} represents the regression of $V$ on $T_b^{1/2}$ as
 \begin{equation}\label{eq:V_PB}
 V = 12.7 (T_b-10\,667)^{1/2} + 432.
 \end{equation}
 The horizontal offset, 10\,667 K, is a parameter that corresponds to the minimum enhancement of the polar brightening. It implies that $T_b-10\,667$ K is an intrinsic enhancement of the polar coronal hole. 
 
 The solar wind velocity appears to be saturated at $V \gtrsim 750$ km/s for $T_b>1.2\times 10^4$ K,  which is the same as for $B>2$ G in figure \ref{fig:V_B}. Therefore, equation \ref{eq:V_PB} is an approximate representation of the saturation feature. The highest velocity from the polar regions has been quite stable for more than 20 years even when the other solar wind parameters (such as the density, temperature, and magnetic field strength) decreased dramatically with the decrease in the solar activity in solar cycle 24 (\opencite{McComas2008GRL}). 
 
 Although the solid line gives a better estimation of $V$ than the dashed line, we should be cautious in using equation (\ref{eq:V_PB}) for space weather forecasting even if the equation can be converted easily to the function of $A$. This is because the equation has not been validated for mid- and low-latitude coronal holes, which significantly affect space weather. 
 The importance of determining these relationships applies to the prediction of the solar wind which can wreak havoc on our electric systems. Therefore, improved prediction algorithms can better prepare us for potential disruptions caused by space weather. 
 
 The relationship $V$\,--\,$A$ is another expression of $V$\,--\,$f$, where $f$ is the magnetic flux expansion rate in the solar corona, defined as $f = A_1/A_0 (R_0/R_1)^2 = B_0/B_1(R_0/R_1)^2$, where $A$, $B$, and $R$ represent the area of the magnetic flux tube, mean magnetic field strength in $A$, and distance from the Sun center, respectively. Suffixes 0 and 1 denote the photosphere and source surface, respectively. 
 A strong anti-correlation was found by \inlinecite{Tokumaru2017SoPh} between $A$ and $f$ (CC = --0.81). It is a natural consequence because the magnetic flux tube originating from a relatively small coronal hole rapidly expands in the corona and balances the magnetic pressure with that from different coronal holes. In addition, identical to the reason of the small coronal hole case, $f$ tends to be larger at the boundary of the coronal hole than the center of the coronal hole. Therefore, the good correlation found in $V$\,--\,$T_b$ through solar cycles is explained using the well-known $V$\,--\,$1/f$ relationship (\opencite{Wang1990ApJ}) in the following scenario.
 During solar minima, the large polar coronal hole (small $f$) persists around the solar poles. In this period, the fast solar wind originates from the polar coronal hole, which occupies most of interplanetary space at mid-latitudes to the solar poles. The slow solar wind originating from the boundary of the polar coronal hole (larger $f$) flows down to lower latitudes around the equator. A shrinking polar coronal hole in the rising phase reduces the fast solar wind region around the solar poles, moving the boundary between fast and slow solar wind to higher latitudes. As a result, the average velocity around the polar region decreases in this period. Then, the slow solar wind originating from a non-polar coronal hole fills the polar regions in interplanetary space during a solar maximum until the reformation of a new polar coronal hole with opposite polarity.

\begin{figure} 
 	\centerline{\includegraphics[width=0.6\textwidth,clip=]{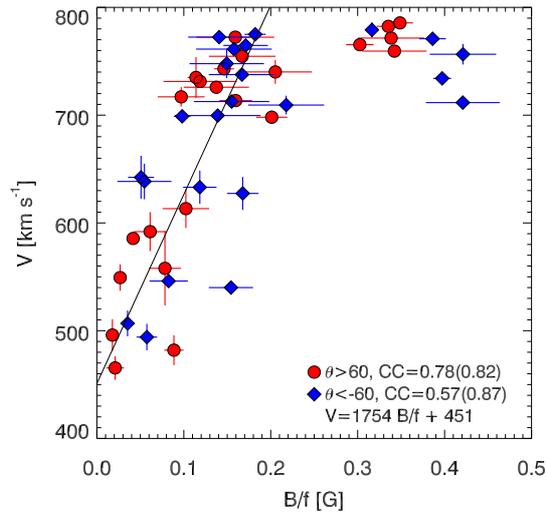}}
 	\caption{
 		Correlation plot between $V$ and $B/f$. The CCs in the two parameters are 0.78 (0.57) for the northern (southern) polar region.  The CCs improve to 0.82 (0.87) merely by using the data points for $B/f<0.25$ G. The black line represents a regression of $V$ on $B/f$, which is derived by using only the data for $B/f<0.25$ G. 
 	} \label{fig:V_Bf}
 \end{figure}

 To predict the solar wind velocity, another useful parameter was proposed by \inlinecite{Fujiki2005AdSpR}. They found a better correlation of $V-B/f$ than that of $V-1/f$ by analyzing the data obtained with the IPS observation KP/NSO, and PFSS calculations around the solar minimum (1995\,--\,1997). Suzuki (2005) theoretically explained this empirical relationship by formulating an energy conservation equation using the parameter, $B/f$, explicitly. \inlinecite{Akiyama2013PASJ} also obtained a better correlation of $V-B/f$ only for the equatorial coronal holes that appeared in 1995\,--\,2005. Figure \ref{fig:V_Bf} shows the correlation plot of $V-B/f$. The CCs are 0.78 (0.57) for the northern (southern) polar regions. The data points appear to be categorized into two groups by the value of $B/f$. $V$ is proportional to $B / f$ in the group with $B / f <0.25$ G, and $V$ is constant in the group with $B / f> 2.5$ G. The CCs improve to 0.82 (0.87) when we use the data points with $B / f <0.25$, which are slightly worse than those with $V-A $ (or $V-1/f$). 
 \inlinecite{Fujiki2015SolPhys} investigated the solar cycle dependence of the $V-B/f$ and $V-1/f$ relationships during 1986\,--\,2009. They found a solar cycle dependence in the regression coefficients of $V-B/f$, which is more pronounced than those of $V-1/f$. Therefore, they concluded that $V-1/f$ is more appropriate for the long-term prediction of the solar wind velocity. This may explain the lower CC observed for $V-B/f$ than that for $V-1/f$ in this study. 
 For the data $B/f<0.25$ G, we obtain a regression line as,
 \begin{equation}
 V=1754\,B/f +451.
 \end{equation}
  %%%%%%%%%%%%%%%%%%%%%%%%%%%%%%%%%%%%%%%%%%%%%%%%%%%%%%
  
%%%%%%%%%%%%%%%%%%%%%%%%%%%%%%%%%%%%%%%%%%%%%%%%%%%%%%

Finally, together with the results obtained by \inlinecite{Akiyama2013PASJ}, who investigated microwave enhancements in the equatorial coronal holes and the solar wind velocity using various parameters such as $A$, $1/f$, and $B/f$, 

a future direction of research can be determined. We focus only on the polar region in this study; however, it is necessary to expand the investigation of the relationships of $V$\,--\,$A$ using a dataset with a higher temporal resolution (one CR) not only in the polar regions but also in lower-latitude regions to further understand the polar brightening and its relation to the solar wind acceleration mechanism.

 \section{Summary \label{sec:summary}}
  We analyzed the long-term variation (1992\,--\,2017) of the polar region observed with NoRH, polar solar wind observed with the IPS, and coronal hole distribution obtained by the PFSS model using the KP/NSO synoptic magnetogram.
 Our main results are summarized below.
 
 The polar brightening strongly correlates with the probability distribution of the open magnetic field region derived by the PFSS model. A remarkably strong correlation (CC = 0.97) is found between the polar brightening and polar coronal hole area derived by PFSS calculations. This result suggests that the polar brightening is strongly coupled to the size of the polar coronal hole. Therefore, the polar brightening is a good indicator of the formation of the polar coronal hole. Even at lower latitudes, brightness enhancements are observed in the radio butterfly map corresponding to the probability distribution of coronal holes. They should be investigated to further understand the brightness enhancement in lower-latitude coronal holes in the future.
 
 The velocity of the polar solar wind is closely correlated with polar brightening. The CCs are 0.91 (0.83) for the northern (southern) polar region. We derived the empirical relationship, $V=12.7 (T_b-10\,667)^{1/2} + 432$ km/s, where $T_b - 10\,667$ K is the intrinsic enhancement of the polar coronal hole. The strong correlation with $V$\,--\,$T_b$ is explained by the $V$\,--\,$A$ relationship because of the strong correlation with $A$\,--\,$T_b$. Considering the anti-correlation with $A$\,--\,$f$, we suggest that the $V$\,--\,$T_b$ is another expression of the Wang--Sheeley relationship in the polar regions.

The velocity of the polar solar wind highly correlates with $B/f$.  The correlation diagram of $V-B/f$ can be divided into two groups using the value around $B/f$=0.25 G at which the solar wind velocity saturates. The group with $B/f<0.25$ G shows linear dependence of $V$ on $B/f$ with CC=0.82 (0.86) for northern and southern polar regions. By using data points only in this group, we obtained empirical relationship as $V=1754 B/f + 451$.

 The brightness temperature on the solar disk possesses a north--south asymmetry. The latitudinal profile of brightness temperature systematically decreases as the northern mid-latitudes are always darker than the southern mid-latitudes during these 25 years. The difference between $40^{\circ} < |\theta| < 50^{\circ}$ is 116 K. If we extrapolate linearly the decline to the pole, the difference will be 233 K at the pole. This difference may cause a systematic deviation of the relationships, $A$\,--\,$T_b$ and $V$\,--\,$T_b$, in the northern and southern polar regions. The reason for the systematic north--south asymmetry is not clear in this work; therefore, it should be investigated using a comprehensive dataset of the Sun in the future. 
 
 %%%%%%%%%%%%%%%%%%%%%%%%%%%%%%%%%%%
%% Acknowledgements
%%%%%%%%%%%%%%%%%%%%%%%%%%%%%%%%%%%
\begin{acks}
NoRH was operated by Nobeyama Solar Radio Observatory (NSRO), National Astronomical Observatory Japan (NAOJ) in 1992\,--\,2015. Since then, it has been operated by the International Consortium for the Continued Operation of Nobeyama Radioheliograph (ICCON). ICCON consists of ISEE/Nagoya University, NAOC, KASI, NICT, and GSFC/NASA. The IPS observations are carried out under the solar-wind program of ISEE/Nagoya University. We are grateful to the KP/NSO for the use of their synoptic magnetogram. 
We also thank COHOWeb/NASA for the use of SWOOPS/Ulysses data. 

This work is partially supported by JSPS KAKENHI, Grant Number 18H01253. 

\end{acks}

%\bibliographystyle{spr-mp-sola}
%\bibliography{ref}  

\end{article} 
\end{document}